\documentclass[amstex,useAMS,epsf,amssymb,usenatbib, twocolumn]{mn2e}
\usepackage{epsfig}
\usepackage{graphicx}
\usepackage{color}
\usepackage{amssymb,amsmath}
\usepackage{multicol}
\usepackage{soul}
\usepackage[normalem]{ulem}
\usepackage{url}
\usepackage{hyperref}
\usepackage{tabularx}

\newcolumntype{Y}{>{\centering\arraybackslash}X}

\newcommand{\gray}{$\gamma$-ray}

\def\aj{AJ}                   
\def\jcap{JCAP}
\def\apj{ApJ}                 
\def\apjl{ApJ}                
\def\apjs{ApJS}               
\def\aap{A\&A}                
\def\aapr{A\&A~Rev.}          
\def\mnras{MNRAS}             
\def\pasp{PASP}               
\def\prd{Phys.~Rev.~D.}

\begin{document}

\title[A geomagnetic filter]{A geomagnetic filter for the \textit{Fermi}-LAT background}

\author[D. A. Prokhorov \& A. Moraghan]{D. A. Prokhorov$^{1}$\thanks{E-mail:phdmitry@gmail.com}
and A. Moraghan$^{2}$
\\
~\\
$^{1}$ School of Physics, Wits University, Private Bag 3, WITS-2050
Johannesburg, South Africa
\\
$^{2}$ Academia Sinica Institute of Astronomy and Astrophysics, PO
Box 23-141, Taipei 106, Taiwan}

\date{Accepted .....
      Received ..... ;
      in original form .....}

\pagerange{\pageref{firstpage}--\pageref{lastpage}} \pubyear{2019}

\maketitle

\begin{abstract}
One of the unsolved questions in gamma-ray astronomy is
whether the extragalactic gamma-ray background is of the
discrete-source origin. To respond to this question, one first needs
to reduce the data by differentiating charged particles
from gamma rays. This procedure is usually performed on the basis of
the detector responses. In this paper, we showed that the
geomagnetic shielding effect at GeV energies can, to some
extent, be used for this purpose for gamma-ray telescopes in a low
Earth orbit. We illustrated this method by applying it to
the \textit{Fermi}-LAT data. To partially decompose the
charge-filtered background, we examined the contribution from
star-forming galaxies by implying a radio/gamma connection in
consideration of next generation radio surveys.
\end{abstract}


\section{Introduction}
\label{sec:intro}

It is recognised that numerous non-thermal sources emit in the broad
frequency range from the radio band to the \gray{} band.
Next-generation, wide field, radio surveys will greatly increase the
number of known non-thermal sources allowing us to make a
significant step towards resolving the extragalactic radio source
background \citep[e.g.,][]{Prandoni2015}. High sensitivity and fine
angular resolution of radio surveys allow studies of sources too
faint for detection in the $\gamma$-ray band. Sources with fluxes
below $\sim$1 millijansky (mJy) at 1.4 GHz contributing to the faint
radio source population include, amongst others, distant
star-forming galaxies (SFGs) which split into populations of
starburst galaxies (SBGs) and normal late-type galaxies (hereafter
normal galaxies) \citep[for a review, see][]{Padovani2016}. The
synchrotron mechanism involving cosmic-ray (CR) electrons spiralling
in magnetic fields is responsible for radio emission from SFGs in
the GHz frequency range, while the decay of neutral pions generated
by the interactions of CR hadrons with the interstellar medium (ISM)
in SFGs is responsible for most of their $\gamma$-ray emission.

Both the $\gamma$-ray background and sub-mJy radio source population
can to a large extent be produced in SFGs.
It is probable that SFGs in the local universe ($z<0.3$)
can be significant contributors to the
$\gamma$-ray background \cite[e.g.,][]{Linden2017} at energies
above 1 GeV and below 10 GeV \citep[see, Fig. 9 from][]{Linden2017}.
Moreover, a significant contribution to the $\gamma$-ray background
at $<10$ GeV can come from SBGs \citep[][]{Thompson2007} and
normal galaxies \citep[][]{Fields2010} at the redshift of peak star
formation $z\approx1-2$.
On the other hand normal SFGs are most likely the dominant
component of the population of sub-mJy radio
sources \citep[e.g.][]{Seymour2008, Bonzini2013}.

Photons with energy below $\sim$10 GeV can travel over cosmological
distances from their sources to the Earth's orbit allowing us to study
physical processes that occur in distant sources
\citep[e.g.,][]{Chen2010}. Unprecedentedly accurate measurements of
the isotropic $\gamma$-ray background are available with the
\textit{Fermi} Large Area Telescope (\textit{Fermi}-LAT)
\cite[][]{Fermi2010, Fermi2015} onboard the \textit{Fermi}
satellite. To determine cumulative $\gamma$-ray emission
coming from sources that are too faint to be individually probed,
one needs to separate $\gamma$ rays from other species of particles
that pass through the \textit{Fermi}-LAT detectors and contribute to
the isotropic particle background. At present, responses of the
\textit{Fermi}-LAT detectors to registered events are used to
identify $\gamma$-ray-like ones.

We developed an algorithm to further remove charged particles from
the \textit{Fermi}-LAT GeV $\gamma$-ray-like background. The
algorithm takes into account that the rate of charged particles
depends on the geomagnetic position of the \textit{Fermi} spacecraft
owing to the fact that the geomagnetic field deflects the low-energy
charged particles back into space. Given \textit{Fermi}'s orbit, the
geomagnetic cut-off for electrons spans from $\simeq$6 to $\simeq$15
GeV. We binned events detected by \textit{Fermi}-LAT into bands in
the McIlwain L parameter which describes magnetically equivalent
positions with respect to an incoming charged particle.
We compared the particle fluxes measured in these bins and argued that
the ones with the lowest measured particle flux are filtered by
the Earth's magnetic field.
Gamma rays corresponding to this McIlwain L bin constitute the charge-filtered
\textit{Fermi}-LAT background. In the \textit{Fermi}-LAT Pass 8
photon data, in particular, for events from the \texttt{SOURCE} and
\texttt{CLEAN} classes, we found evidence for both

(i) the geomagnetic shielding and

(ii) the east-west effect\\
which are expected only for charged particles.

We used the values of the WP8CTAllProb and WP8CTCalTkrProb
parameters recorded in the extended \textit{Fermi}-LAT files each
corresponding to a probability that the given particle is a $\gamma$
ray (versus a cosmic ray) as estimated on the basis of the LAT
tracker (TKR) and calorimeter (CAL) variables both with or without
accounting for the anti-coincidence detector (ACD), respectively. We
computed their mean probability in every McIlwain L bin and in every
energy band and established that the low mean probabilities of
WP8CTCalTkrProb correspond to those bins and bands in which the
fluxes reveal evidence for the effects (i) and (ii). It shows that
the geomagnetic field serves as a natural filter of charged
particles.

We compared the measured charge-filtered background flux
with the $\gamma$-ray fluxes derived on the basis of the radio
source counts to check if various populations of SFGs can provide an
explanation for the extragalactic $\gamma$-ray background. We
calculated the fluxes with $\gamma$-ray energies of 3 GeV
to 8 GeV from the SFGs in the framework of scenarios in
which (1) their radio source counts are described by the model based
on semi-empirical simulations by \citet[][]{Wilman2008} and (2) both
the SFGs in the local universe ($z<0.3$) and distant SFGs (including
those at the redshift of peak star formation) are similar in their
observational radio and $\gamma$-ray properties to the
nearby, paradigmatic SFGs.
Under these assumptions the total contribution produced in SFGs is between 10\% and 25\% of
the total unresolved background flux in this energy band.
Thus, if distant SFGs are dominant contributors then the physical properties of galaxies
significantly evolved between the redshift of peak star formation and today, or the $\mu$Jy radio
population of SFGs is larger than simulated, an assumption that may be confirmed through deep radio surveys.

\section{Fermi-LAT observations and analysis}
\label{sec:analysis}

\subsection{Filters of astroparticles}

The \textit{Fermi}-LAT is a wide-field-of-view, pair-conversion,
$\gamma$-ray telescope with a TKR which consists of tungsten
converter layers and silicon tracker planes and with a CAL which
contains an array of scintillating crystals. A segmented ACD made of
plastic scintillator tiles is the outermost detector layer in the
telescope and sensitive to incident CR charged particles, but not to
$\gamma$ rays. Being a pair-conversion $\gamma$-ray telescope, the
\textit{Fermi}-LAT is also an electron-positron detector. The
\textit{Fermi}-LAT has been scanning the sky continuously since
August 2008 providing all-sky coverage every $\sim$3 hours (2
spacecraft orbits).

The rate of CR background events passing through the
\textit{Fermi}-LAT is 2-4 kHz and is dominated by CR protons with
smaller helium, electron, and positron components
\citep[][]{Atwood2009}. With a primary goal of detecting $\gamma$
rays, a set of particle discriminators are applied to events
collected by \textit{Fermi}-LAT in order to reduce the residual
background of misclassified charged particles to the rate of $<$1
Hz, which is similar to that of the extragalactic $\gamma$-ray
background. These discriminators are based on the response of the
\textit{Fermi}-LAT instrument to detected events including a signal
from the ACD in conjunction with the tracks found in the TKR and a
match of the shower profile in three dimensions in both the TKR and
CAL.

The \textit{Fermi}-LAT instrument is on board the \textit{Fermi} spacecraft which is in a nearly
circular orbit of 565 km altitude with an orbital inclination of 25.6$^{\circ}$.
The orbital parameters of the \textit{Fermi} satellite define the geomagnetic
latitude range over which the data are accumulated and also the range of geomagnetic
cut-off values \citep[e.g.][]{electrons2010, electrons2012}.
At energies above a few GeV, the LAT has a point-spread function (PSF) narrower than
1.0$^{\circ}$ \citep[][]{Fermi2013, PM2016} allowing us to remove the contribution of known
$\gamma$-ray point sources from the total observed signal and therefore to use a significant
fraction of the sky to evaluate the unresolved $\gamma$-ray background.

\begin{figure*}
    \centering
    \includegraphics[width=0.75\textwidth]{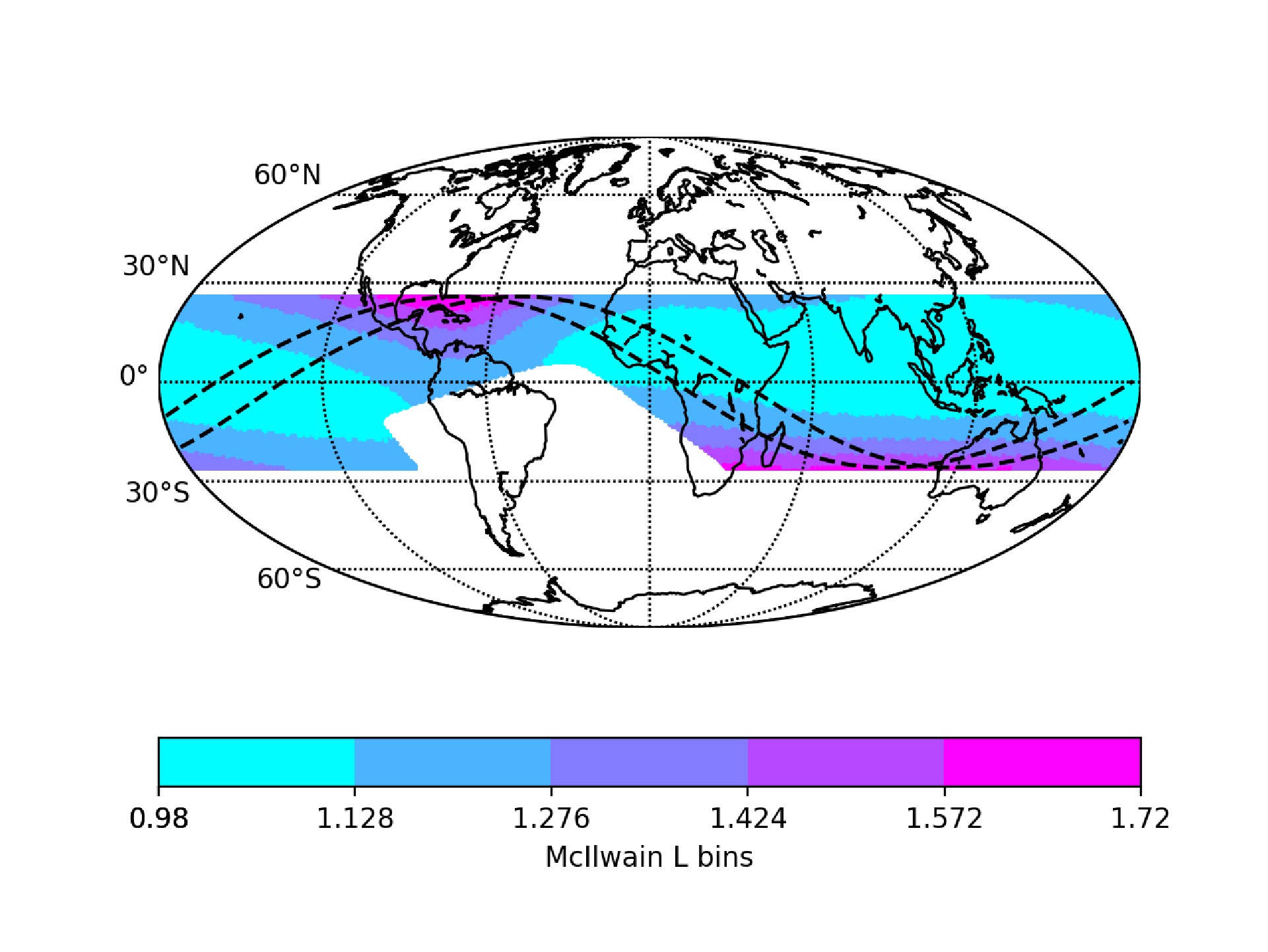}
    \caption{Figure shows the McIlwain L map and two arbitrary Fermi's orbits intersecting
    the regions corresponding to the McIlwain L bin 1.572$<$L$<$1.720.
    This figure can be seen in colour in the online version.}
    \label{Fig1}
\end{figure*}

Magnetically equivalent positions with respect to an incoming
charged particle correspond to a set of the geomagnetic field lines
crossing the Earth's magnetic equator at a number of Earth-radii
equal to the McIlwain L-parameter value. The inclination of the
\textit{Fermi} satellite's orbit fixes the range of McIlwain L
values to 0.98-1.72. Taking into account that the geomagnetic energy
cut-off shifts towards lower energies with increasing McIlwain L
parameter, the contribution of charged particles above a
given energy increases with the McIlwain L-parameter
\citep[for the sake of illustration, see Figure 13
from][]{electrons2010}. The shielding effect of the geomagnetic
field is therefore stronger (weaker) for smaller (larger) values of
McIlwain L-parameter. We show that the motion of the \textit{Fermi}
satellite orbiting the Earth provides us with a tool for filtering
charged particles hitting the \textit{Fermi}-LAT's detectors during
the time intervals with a high charged-particle rate. Two
arbitrary \textit{Fermi's} orbits shown in Figure
\ref{Fig1} illustrate the fact that the \textit{Fermi}
spacecraft passes the regions corresponding to the highest
McIlwain L bin 1.572$<$L$<$1.720.

Moreover, the Earth's magnetic field, by means of the east-west effect,
allows us to distinguish if
these charged particles are CR electrons or positrons. This is
possible owing to the fact that CR electrons arrive from the
eastward direction while CR positrons arrive from the westward
direction resulting in an asymmetry in the azimuthal distribution.

\subsection{Data reduction}

We downloaded the \textit{Fermi}-LAT Pass 8 (P8R2) photon data from
the \textit{Fermi} Science Support Center FTP
Server\footnote{\texttt{http://heasarc.gsfc.nasa.gov/FTP/fermi/data/lat/weekly/}}.
The data accumulated during each week of the \textit{Fermi}'s
science mission are contained in weekly data files (called photon
files), while the \textit{Fermi} spacecraft location/orientation
information for each mission week is contained in the spacecraft
files. For the data analysis, we used the \texttt{FERMI SCIENCE
TOOLS}
package\footnote{\texttt{https://fermi.gsfc.nasa.gov/ssc/data/analysis/software/}}.
We selected the data spanning 9.4 yr (August 7, 2008\footnote{For
the sake of convenience we ignore the short observing week between
August 4, 2008 and August 6, 2008.} - January 4, 2018) with energies
between 3 GeV and 8 GeV at which the contribution from SFGs to the
$\gamma$-ray background is expected to be dominant. To avoid
contamination from the $\gamma$-ray-bright Earth's limb, we removed
all events with zenith angle larger than 90$^{\circ}$. To exclude
time intervals when some spacecraft event has negatively affected
the quality of the \textit{Fermi}-LAT data, we applied the
recommended quality cuts, \texttt{DATA\_QUAL>0 \&\& LAT\_CONFIG==1}.
We also applied the cut on McIlwain L-parameter values which are
given in the spacecraft location information. To make the statistics
larger we selected events converted anywhere in the TKR. We binned
the data using the \texttt{HEALPix} package \citep[][]{Gorski2005}
into a map of resolution $N_{\mathrm{side}}=256$ in Galactic
coordinates with ``RING'' pixel ordering. With these settings, the
total number of pixels is equal to 786432 (that is $12
N^2_{\mathrm{side}}$) and the area of each pixel is
$1.598\times10^{-5}$ sr.
The resolution of the map is chosen according to the size of the \textit{Fermi}-LAT
PSF in the 3 GeV to 8 GeV energy range.

To analyse the variations of a background flux with the geomagnetic
position, we binned \textit{Fermi}-LAT events in five equidistant McIlwain L
bins between 0.98 and 1.72.
For each McIlwain L bin, we further binned the \textit{Fermi}-LAT events in
three energy bands, namely, 3-4 GeV, 4-6 GeV, and 6-8 GeV. The choice of
energy bands is based on the fact that CR electron rates
measured with \textit{Fermi}-LAT in the lowest and highest McIlwain L bins
are affected by non-shielded electrons above $\simeq$7 GeV and
$\simeq$4 GeV, respectively \citep[see Fig. 13 from][]{electrons2010}.
Since the \textit{Fermi} spacecraft spends a fraction of time during every
orbit within each of these McIlwain L bins, to compute the fluxes
corresponding to each McIlwain L bin we recorded the start and end
times at which the spacecraft enters and leaves positions within
McIlwain L bins. For each \texttt{HEALPix} pixel and each time
interval, we recorded the number of photons and computed the
exposure using the \texttt{gtbin} and \texttt{gtexpcube2} routines.
After that we added the values corresponding to the same McIlwain L bin
together and obtained the combined count and exposure maps for
the five McIlwain L bins.

In the Pass 8 data release, $\gamma$-ray-like event classes with a low residual
charged-particle background rate include \texttt{P8R2\_SOURCE},
\texttt{P8R2\_CLEAN}, \texttt{P8R2\_ULTRACLEAN}, and
\texttt{P8R2\_ULTRACLEANVETO}.
The \texttt{P8R2\_SOURCE} class is recommended for most of the
\textit{Fermi}-LAT analyses. Above 3 GeV the \texttt{P8R2\_CLEAN}
class has a lower background rate than \texttt{P8R2\_SOURCE}, but is
identical to \texttt{P8R2\_SOURCE} below 3 GeV. The
\texttt{P8R2\_ULTRACLEANVETO} class is the cleanest and is
recommended for the studying of CR-induced systematics, while
\texttt{P8R2\_ULTRACLEAN} has a background rate between the rates
for the \texttt{P8R2\_CLEAN} and \texttt{P8R2\_ULTRACLEANVETO}
classes. With the goal of identifying the geomagnetic shielding
effect of charged particles in the satellite's low Earth orbit,
we compared the results obtained for the
\texttt{P8R2\_SOURCE} and \texttt{P8R2\_ULTRACLEANVETO} classes.
To quantify the improvement on the separation
efficiency of $\gamma$ rays from charged particles, we also
show the results for the \texttt{P8R2\_CLEAN} class.

In order to evaluate the unresolved $\gamma$-ray background, one
needs to mask the known sources of $\gamma$-ray emission. We masked
the region between the Galactic latitudes of, $-45^{\circ}<b<+45^{\circ}$,
to reduce the contribution of bright diffuse $\gamma$-ray emission from
the Galactic plane. We also repeated the analysis applying a broader mask,
$-60^{\circ}<b<+60^{\circ}$,
to test the uncertainties caused by an intermediate-latitude Galactic
foreground emission. We used the \textit{Fermi} LAT 4-year point
source catalogue \citep[][]{LATcat}, gll\_psc\_v16.fit, to list
coordinates of the known $\gamma$-ray point sources and to mask
regions of 2$^{\circ}$ radius centred on the positions of these
sources by applying the \texttt{healpy.query\_disc} tool. We also
masked extended $\gamma$-ray sources, the northern and southern
Fermi bubbles and the Large and Small Magellanic Clouds, by
excluding \texttt{HEALPix} pixels within the disks of 32$^{\circ}$,
32$^{\circ}$, 7$^{\circ}$, and 5.5$^{\circ}$ radius placed at the
centres of these extended sources at Galactic coordinates
(Longitude, Latitude) of (358.90$^{\circ}$, 25.0$^{\circ}$),
(355.4$^{\circ}$, -29.0$^{\circ}$), (278.75$^{\circ}$,
-32.8$^{\circ}$), and (302.20$^{\circ}$, -44.38$^{\circ}$)
respectively.

In order to compute the unresolved background fluxes in Section \ref{subsec:geo},
we subtracted a pedestal level of high-latitude Galactic foreground emission
from the total flux derived after masking the aforementioned point and extended
$\gamma$-ray sources. We calculated a pedestal level of the high-latitude Galactic flux
by applying the \texttt{gtmodel} routine to the Galactic emission template, gll\_iem\_v06.fits,
provided by the \textit{Fermi}-LAT collaboration and selecting the north and south
Galactic caps with a 45$^{\circ}$ (or 30$^{\circ}$) radius as our regions of interest.
The obtained $\gamma$-ray fluxes of high-latitude Galactic emission ${\mid}b\mid>45^{\circ}$ (${\mid}b\mid>60^{\circ}$)
are $4.86\times10^{-8}$, $4.00\times10^{-8}$ and $1.57\times10^{-8}$ ($4.55\times10^{-8}$, $3.75\times10^{-8}$
and $1.47\times10^{-8}$) ph cm$^{-2}$ s$^{-1}$ sr$^{-1}$ for the energy bands of 3-4 GeV,
4-6 GeV and 6-8 GeV, respectively.

\section{Fermi-LAT background}

\subsection{Geomagnetic shielding}
\label{subsec:geo}

We computed the background fluxes in the McIlwain L bins and in the
energy bands to check if charged particles are present amongst the
detected \textit{Fermi}-LAT events of the three event classes. The
computed fluxes are shown in Table \ref{tab1}. If all particles were
$\gamma$-ray photons, one would expect that the fluxes within a
given event class and a given energy band are the same for all
McIlwain L bins. However, we found that the fluxes for the
\texttt{SOURCE} event class and also for the \texttt{CLEAN} event
class in the energy bands of 4-6 GeV and 6-8 GeV for large values of
McIlwain L-parameter are by a factor of about 2.5-4.0 and 2.3-3.0
higher (see the values shown in bold in Table \ref{tab1}) than the
fluxes in the 0.980-1.128 McIlwain L bin for these event classes,
respectively, in these two energy bands. The statistical
errors on the fluxes in the bins with excesses calculated by the
square root of the count number divided by the exposure are
$1.3-2.7\%$ and $2.1-5.2\%$ for the regions of
${\mid}b\mid>45^{\circ}$ and ${\mid}b\mid>60^{\circ}$, respectively.
The observed excesses in flux are statistically significant.
The behaviour of the fluxes with McIlwain
L parameter is similar to that which is expected if the shielding
effect due to the geomagnetic field plays a significant role in the
formation of the observed fluxes.

\begin{table*}
\centering \caption{The $\gamma$-ray fluxes for various McIlwain L
bins and energy bands obtained after masking events from the
Galactic plane, $-45^{\circ}<b<+45^{\circ}$ (or
$-60^{\circ}<b<+60^{\circ}$). Fields in bold correspond to flux excesses
that can be attributed to charged particles (See Section 3.2.1). }
\begin{tabular}{ | c | c | c | c | c |}
\hline
& \multicolumn{4}{r}{Fluxes, $10^{-8}$ cm$^{-2}$ s$^{-1}$ sr$^{-1}$} \\
\cline{3-5}
Event class & McIlwain L bin & 3-4 GeV & 4-6 GeV & 6-8 GeV\\
\hline
\texttt{SOURCE} & 0.980-1.128 & $3.79 (3.25)$ & $3.28 (2.86)$ & $1.59 (1.46)$ \\
\texttt{SOURCE} & 1.128-1.276 & $3.60 (3.20)$ & $3.60 (3.22)$ & $1.83 (1.71)$ \\
\texttt{SOURCE} & 1.276-1.424 & $3.80 (3.49)$ & $4.15 (3.80)$ & $2.73 (2.61)$ \\
\texttt{SOURCE} & 1.424-1.572 & $4.63 (4.07)$ & $5.35 (4.90)$ & $\textbf{4.58 (4.32)}$ \\
\texttt{SOURCE} & 1.572-1.720 & $5.30 (5.38)$ & $\textbf{8.22 (7.75)}$ & $\textbf{5.98 (5.87)}$ \\
\texttt{SOURCE} & Combined flux   & $3.84 (3.41)$ & $3.80 (3.42)$ & $2.18 (2.07)$ \\
\hline
\texttt{CLEAN} & 0.980-1.128 & $3.68 (3.13)$ & $3.16 (2.74)$ & $1.49 (1.35)$ \\
\texttt{CLEAN} & 1.128-1.276 & $3.51 (3.12)$ & $3.45 (3.08)$ & $1.69 (1.56)$ \\
\texttt{CLEAN} & 1.276-1.424 & $3.71 (3.41)$ & $3.94 (3.57)$ & $2.28 (2.13)$ \\
\texttt{CLEAN} & 1.424-1.572 & $4.53 (3.99)$ & $4.89 (4.42)$ & $\textbf{3.50 (3.45)}$ \\
\texttt{CLEAN} & 1.572-1.720 & $5.15 (5.28)$ & $\textbf{7.38 (6.95)}$ & $\textbf{4.18 (3.99)}$ \\
\texttt{CLEAN} & Combined flux   & $3.74 (3.31)$ & $3.60 (3.22)$ & $1.90 (1.78)$ \\
\hline
\texttt{ULTRACLEANVETO} & 0.980-1.128 & $2.72 (2.24)$ & $2.68 (2.29)$ & $1.38 (1.24)$ \\
\texttt{ULTRACLEANVETO} & 1.128-1.276 & $3.04 (2.55)$ & $3.13 (2.68)$ & $1.46 (1.36)$ \\
\texttt{ULTRACLEANVETO} & 1.276-1.424 & $3.04 (2.72)$ & $3.30 (2.97)$ & $1.50 (1.27)$ \\
\texttt{ULTRACLEANVETO} & 1.424-1.572 & $3.55 (3.10)$ & $3.57 (3.16)$ & $\textit{1.75 (1.60)}$ \\
\texttt{ULTRACLEANVETO} & 1.572-1.720 & $3.93 (3.99)$ & $\textit{4.27 (4.10)}$ & $\textit{1.60 (1.46)}$\\
\texttt{ULTRACLEANVETO} & Combined flux   & $2.94 (2.52)$ & $3.00 (2.63)$ & $1.46 (1.32)$\\
\hline
\end{tabular}
\label{tab1}
\end{table*}

In the presence of a shielding effect, one expects that

(i) if charged particles are present in these data then there are no significant
flux variations in the energy band of 3-4 GeV \citep[see Fig. 13 from][]{electrons2010};

(ii) if charged particles are present in these data then flux
variations with McIlwain L parameter should occur in the
energy bands of 4-6 GeV and to a great degree in the 6-8 GeV band
\citep[see Fig. 13 from][]{electrons2010}.

Both the expectations (i) and (ii) are in agreement with the results shown in Table \ref{tab1}.

We found that variations in the measured fluxes with McIlwain parameter for the \texttt{ULTRACLEANVETO}
class is less pronounced. The fields in Table \ref{tab1} shown in italics correspond
to those contaminated by charged particles for the \texttt{SOURCE} and
\texttt{CLEAN} event classes.

We computed the exposure-weighted background fluxes for the
\texttt{SOURCE} and \texttt{CLEAN} event classes in the three energy
bands by means of a combined analysis of McIlwain L bins. We checked
and found that the numbers of detected events in the bins with
smaller values of McIlwain L-parameter are higher and their
contribution to the combined flux is dominant. However, we also
found that the relative differences between the combined fluxes and
the fluxes in the lowest McIlwain L bin (that is with the lowest
charged particle rate) are as high as $(29\pm4_{\mathrm{sys}})$\%
and $(24\pm3_{\mathrm{sys}})$\% for the 6-8 GeV energy band for the
\texttt{SOURCE} and \texttt{CLEAN} event classes (and as high as
$(15\pm2_{\mathrm{sys}})$\% and $(16\pm2_{\mathrm{sys}})$\% for the
4-6 GeV energy band for these two classes). The statistical errors
on these differences are significantly smaller than the systematic
ones owing to the fact that the number of events, $N$, in the lowest
McIlwain L bin is the largest and the value of
$\left({\Delta}N_{\mathrm{stat}}/N\right)\simeq1/\sqrt{N}$ for these
bins is between 0.011 and 0.017, while the systematic uncertainty of
the effective area for the \texttt{SOURCE} class events is estimated
to be $\left({\Delta}N_{\mathrm{sys}}/N\right)\simeq0.10$. The
events recorded during the time intervals while the \textit{Fermi}
spacecraft's geomagnetic position corresponded to the lowest
McIlwain L-parameter bin constitute the charge-filtered
\textit{Fermi}-LAT background for the \texttt{SOURCE} and
\texttt{CLEAN} classes. The fluxes corresponding to these
charge-filtered \textit{Fermi}-LAT backgrounds are close to (though
somewhat higher than) those obtained for the \texttt{ULTRACLEANVETO}
class.

We also checked that if one adds events from the \texttt{CLEAN} class
in the McIlwain L bin of 0.980-1.128 to events from the \texttt{ULTRACLEANVETO}
class in the McIlwain L bins with the L parameter above 1.128 then
the event sample will increase by 12-14 per cent compared to the \texttt{ULTRACLEANVETO}
sample in the energy bands of 4-6 GeV and 6-8 GeV. The joint sample will be filtered
by both the LAT detectors and Earth's magnetic field.

\subsection{Two tests on the origin of excesses}

To confirm the charged-particle origin of excesses in the unresolved background flux
for the \texttt{SOURCE} and \texttt{CLEAN} event classes, we performed two tests
based on (1) a search for the east-west effect in the data and (2) estimators which
give a probability of particle events to be $\gamma$ rays. For these tests we masked
the regions between the Galactic latitudes of $-45^{\circ}<b<+45^{\circ}$
($-60^{\circ}<b<+60^{\circ}$) while leaving the 3FGL sources with higher Galactic latitudes
unmasked because of computational time requirements for sorting events individually.

\subsubsection{East-west effect}

The Earth's magnetic field introduces an asymmetry between the flux
of the incoming CRs from the east and from the west. The positively
charged particles, including positrons (and protons), come from the
west owing to the fact that they are deflected towards the east
by the Earth's magnetic field, while the negatively charged
particles, electrons, come from the east. Thus, if the telescope
measures the flux ratio of CR particles coming from the east to CR
particles coming from the west above unity in the dataset, then it
will indicate the presence of primary CR electrons. To select the
events coming from the east or from the west it is necessary to
include a selection criterion based on the reconstructed arrival
direction of the event in the detector.
The angle, \texttt{EARTH\_AZIMUTH\_ANGLE}, of the reconstructed event
direction with respect to North (line from spacecraft origin to
north celestial pole) as projected onto a plane normal to the zenith
is recorded in the photon file. This angle is measured in degrees
east of north, such that 90 degrees indicates that the event
originated from the
west\footnote{https://fermi.gsfc.nasa.gov/ssc/data/analysis/documentation/\\
Cicerone/Cicerone\_Data/LAT\_Data\_Columns.html}. The azimuthal flux
ratio of \textit{Fermi}-LAT events coming from the east to those
coming from the west in $45^{\circ}$ intervals normalised to that in
the McIlwain L bin, 0.980-1.128, is shown in Table \ref{tab2}. We
found that the computed ratio significantly exceeds the expectation
(that is $\simeq$ 1) for the McIlwain L bins with background
excesses and for the \texttt{SOURCE} and \texttt{CLEAN} classes
and is $(1.92\pm0.13_{\mathrm{stat}})$,
$(1.82\pm0.12_{\mathrm{stat}})$, $(2.17\pm0.12_{\mathrm{stat}})$,
$(1.61\pm0.13_{\mathrm{stat}})$, $(1.85\pm0.11_{\mathrm{stat}})$,
and $(1.45\pm0.13_{\mathrm{stat}})$ in the corresponding bins.

If the variation in the normalised azimuthal ratio with McIlwain L
parameter is due to the presence of charged particles, then it is
possible to estimate their contribution to the total flux in each
McIlwain L bin assuming that the reconstructed direction of
the photon-like event corresponds to that of the misclassified CR.
Under this assumption we found using Table \ref{tab2} that 30-45\% of
the fluxes for the \texttt{SOURCE} and \texttt{CLEAN} event classes
in the McIlwain L bins of 1.424-1.572
and 1.572-1.720 and the 6-8 GeV energy band and in the McIlwain L
bin of 1.572-1.720 and the 4-6 GeV energy band (shown in bold in
Table \ref{tab1}) can be attributed to charged particles. We checked
and found that the presence of these particles in the data can, to a
large extent, explain the flux excesses shown in Table \ref{tab1}.
Thus, the relative difference between the 4-6 GeV $\gamma$-ray fluxes
in the McIlwain L bins, 1.424-1.572 and 1.572-1.720,
decreases from 58\% to 12\% (if the presence of charged particles in the latter bin is
assumed) and the relative difference between the 6-8 GeV $\gamma$-ray fluxes
in the McIlwain L bins, 1.276-1.424 and 1.424-1.572, decreases from 66\% to 13\%.

\subsubsection{WP8CTCalTkrProb parameter}

While both the geomagnetic shielding and east-west effect are natural tools
to check if charged particles are present in the data, the \textit{Fermi}-LAT
instrument has possibilities for tracking and calorimetry allowing us to distinguish
$\gamma$ rays from other particles even without accounting for the ACD.
The \textit{Fermi}-LAT extended file selection is a superset of the photon file selection.
These files also contain additional event-level quantities that are not available in the photon files.
Amongst these quantities there are probabilities that the particle is a cosmic ray versus a $\gamma$ ray
(0=CR-like, 1=$\gamma$ ray-like). Thus, we used two quantities, WP8CTCalTkrProb and WP8CTAllProb, to
check the reliability that the background excesses are of charged-particle origin.
There are four possible cases including

(i) if both the quantities show low mean probabilities of particle
events to be $\gamma$ rays in the McIlwain L bins with background
excesses then it will indicate the charged particle origin of excesses;

(ii) if only the WP8CTCalTkrProb quantity shows low mean probabilities of particle events
to be $\gamma$ rays in the McIlwain L bins with background excesses then it will imply that some of
these particles are charged ones, but do not hit ACD filter tiles;

(iii) if only the WP8CTAllProb quantity shows low mean probabilities of particle events
to be $\gamma$ rays in the McIlwain L bins with background excesses then it will imply
that tracking and calorimetry have insufficient power for identifying
some of these events;

(iv) if both the quantities do not show low mean probabilities of particle events to
be $\gamma$ rays in the McIlwain L bins with background excesses then it will imply that an
alternative scenario for background excesses is needed.

We chose the mean probability within each of the McIlwain L bins and
each of the energy bands to characterise the distribution
of probabilities. We did not find any variation of the WP8CTAllProb
mean probability with the McIlwain L parameter within each of the
three energy bands for the \texttt{SOURCE}, \texttt{CLEAN} and
\texttt{ULTRACLEANVETO} event classes. We found that the
WP8CTCalTkrProb mean probability is the lowest in the McIlwain L
bins with excesses in background fluxes (see Table \ref{tab3}).
Thus, the case (ii) gained considerable support. Possible
interpretations of this result include leakage of CR electrons
through the ACD ribbons and pileup effects in the ACD.

\section{Radio/$\gamma$-ray connection in SFGs}

\begin{table*}
\centering \caption{The fluxes of populations of SBGs and normal
galaxies calculated in the framework of two models and
compared to that of the observed $\gamma$-ray background.
All the differential fluxes are expressed as
$E\mathrm{d}F/\mathrm{d}E$ (GeV
cm$^{-2}$s$^{-1}$GeV$^{-1}$sr$^{-1}$).}
\begin{tabular}{ | c | c | c | c | c | c |}
\hline
Model & Galaxies & Prototype & Flux at 3.5 GeV & Flux at 5 GeV & Flux at 7 GeV \\
\hline
Shell & SBG & NGC 253 & $5.6\times10^{-9}$ & $3.7\times10^{-9}$ & $2.5\times10^{-9}$ \\
         & SBG & Arp 220 & $1.6\times10^{-8}$ & $1.1\times10^{-10}$ & $7.3\times10^{-9}$ \\
         & normal & Andromeda & $8.3\times10^{-9}$ & $5.1\times10^{-9}$ & $3.2\times10^{-9}$ \\
         & normal & Milky Way & $3.0\times10^{-9}$ & $1.6\times10^{-9}$ & $9.5\times10^{-10}$ \\
\hline
Uniform & SBG & NGC 253 & $6.7\times10^{-9}$ & $4.4\times10^{-9}$ & $3.0\times10^{-9}$ \\
         & SBG & Arp 220 & $1.4\times10^{-8}$ & $9.5\times10^{-9}$ & $6.2\times10^{-9}$ \\
         & normal & Andromeda & $9.7\times10^{-9}$ & $6.0\times10^{-9}$ & $3.7\times10^{-9}$ \\
         & normal & Milky Way & $3.7\times10^{-9}$ & $2.0\times10^{-9}$ & $1.2\times10^{-9}$ \\
\hline
Observed & -- & -- & $8.6\times10^{-8}$ & $6.2\times10^{-8}$ & $4.5\times10^{-8}$ \\
\hline
\end{tabular}
\label{tab5}
\end{table*}

Although our algorithm shows that charged particles contribute to the
\textit{Fermi}-LAT backgrounds of photon-like \texttt{SOURCE} and
\texttt{CLEAN} class events, we clarified that these backgrounds can be
filtered by selecting events only from the lowest McIlwain L bin.
The performed analysis also shows that the \texttt{ULTRACLEANVETO} class
is significantly less affected by these charged particles.
Taking this result into account, we examined the radio/$\gamma$-ray connection
in SFGs with the goal of searching for signs of evolutionary effects.

\subsection{CR propagation regimes}

A discrete-source origin of the isotropic $\gamma$-ray background
was proposed by \citet{Strong1976}. In their paper,
the authors considered this possibility for estimating the
contribution of normal galaxies to the unresolved $\gamma$-ray
background and found that about 4 per cent of the background can be
produced by these sources. One of their methods is based on the
assumption of the same $\gamma$-ray-to-radio flux ratio for all
normal galaxies. \citet{Thompson2007} showed that a similar
assumption is valid for SBGs if these galaxies are CR particle
calorimeters\footnote{A CR calorimeter is a galaxy in which
CR losses dominate over the other losses, including escape.}
and radio-emitting electrons and positrons are created
in the $pp$ collisions. The argument of CR (electron and proton)
calorimetry by \citet[][]{Thompson2007} is particularly compelling
for SBGs, where the energy loss times of CRs are significantly
shorter than those in normal star-forming galaxies. The modelled
$\gamma$-ray background coming from SBGs depends sensitively on the
fraction of star formation that occurs in the CR proton calorimeter
limit at every redshift. Since at high redshift a much larger
fraction of star formation occurs in high surface density systems
which are most likely proton calorimeters, the significant
contribution from high redshift SBGs to the $\gamma$-ray background
was proposed. Contrary to long-standing expectations,
\citet[][]{Fields2010} found that the contribution of numerous but
individually faint normal galaxies to the $\gamma$-ray background is
significant (a) if the redshift history of cosmic star
formation is taken into account and (b) if the ratio of a galaxy's
volume-averaged CR proton flux to the cosmic star formation rate is
constant for all normal galaxies. The source of $\gamma$ rays
and radio-emitting secondary CR electrons in the SBGs is collisions
of CR protons with the ISM. On the other hand, most of the radio-emitting
CR electrons in normal galaxies are primary and originated along with
the CR protons in supernova remnants. In both cases, the source
function of CR electrons is linearly proportional to that of CR protons
resulting in the radio/$\gamma$-ray connection for SFGs, but
the coefficients of proportionality are different and depend on a
CR propagation regime.

The effect due to a transition from the regime of CR proton escape
to the calorimetry regime for SFGs due to the increase in gas
density with redshift is likely to occur. Relating the $\gamma$-ray
background to the radio background (in particular to the radio
source counts) gives us a suitable method
\citep[e.g.,][]{Strong1976} to examine the significance of this
effect in the evolution of the observational properties of galaxies.

\subsection{Prototype SFGs in the radio and $\gamma$-ray bands}

We considered $\gamma$-ray detected SFGs, including nearby SBGs NGC
253 and M82 \citep[for measured fluxes see][]{latsbg,
kapinska2017, Williams2010}, a luminous infrared galaxy NGC 1068
\citep[][]{Lenain2010}, an ultraluminous infrared galaxy Arp 220
\citep[e.g.][]{Griffin2016, Varenius2016} and normal galaxies, the
Milky Way \citep[e.g.,][]{Ackermann2012} and the Andromeda galaxy
\citep[e.g.,][]{Dennison1975, Ackermann2017} to explore the
connection between their radio and $\gamma$-ray emission.
Starburst galaxies undergo an exceptionally high rate of
star formation as compared to the rate of star formation in the
Milky Way. For these prototype galaxies the star-formation rate
varies from 0.4 $M_{\odot}\mathrm{yr}^{-1}$ in the Andromeda galaxy
\citep[][]{Rahmani2016} to 220 $M_{\odot}\mathrm{yr}^{-1}$ in
Arp 220 \citep[][]{Varenius2016}, which is the most active
star-forming galaxy in the local universe. The other galaxies are
included in order to fill the broad range of star-formation rates.
Given that the star-formation rate increases as the insterstellar
gas becomes denser \citep[][]{Schmidt1959}, these galaxies are expected
to behave in various CR propagation regimes. For this investigation
we also took $\gamma$-ray fluxes from \citet[][]{LATcat} and radio fluxes
compiled in \citet[][]{Thompson2007}. Using the measured radio and
$\gamma$-ray band fluxes, we computed the coefficient of
proportionality between the 1.4 GHz radio continuum flux of 1 Jy and
the $\gamma$-ray flux, $E\mathrm{d}F/\mathrm{d}E$, at energies of
3.5 GeV, 5 GeV and 7 GeV in GeV cm$^{-2}$s$^{-1}$GeV$^{-1}$ for each
of these galaxies (see Table \ref{tab4}). We found that the
$\gamma$-ray-to-radio flux ratios are the lowest for normal
galaxies, the Milky Way and the Andromeda galaxy, and are the
highest for Arp 220 and NGC 253. Table \ref{tab4} also shows that
the variation in $\gamma$-ray-to-radio flux ratios for these SFGs is
about a factor of 40 (or 10 if Arp 220 is excluded). This variation
is likely to be related to the fact that the CR proton losses are
dominated by escape in normal galaxies and by pion production in
SBGs. Given that the radio emission from Arp 220 is dominated by two
bright nuclei in the GHz band while its diffuse optically thin
synchrotron flux is 3.3 times less than the total flux at 1 GHz
\cite[see Fig. 3 from][]{Varenius2016}, the existence of SBGs with
the $\gamma$-ray-to-radio flux ratio higher than that for Arp 220 is
plausible. Although the $\gamma$-ray detection of nearby SFGs became
possible only with \textit{Fermi}-LAT, their radio fluxes are of the
order of 1 Jy (see Table \ref{tab4}) and exceed the current limit of
sensitivity in the radio band by several orders of magnitude.

Given that SFGs with far infrared (FIR) luminosities between
$10^{10.5} L_{\odot}$ and $10^{12.5} L_{\odot}$ in the local
universe ($z<0.3$) can possibly be significant contributors to the
isotropic extragalactic $\gamma$-ray background at energies above 1
GeV and below 10 GeV \citep[][]{Linden2017}, it is worth noting that
this luminosity range includes the SBGs such as NGC 253 at the low
FIR luminosity limit and Arp 220 at the high FIR luminosity limit.
At $z=0.3$ the luminosity distance is 1566 Mpc, and if galaxies
identical to NGC 253 and Arp 220 are at this redshift, their radio
fluxes at 1.4 GHz will be 50 $\mu$Jy and 0.6 mJy, respectively. To
perform these calculations we took the spectral index for NGC 253
from \citet[][]{kapinska2017} and assumed that the radio spectrum of
Arp 220 is roughly flat in the GHz frequency band
\citep[][]{Varenius2016}. Source counts for SFGs generated from the
simulation catalogue by \citet[][]{Wilman2008} are given for radio
fluxes at 1.4 GHz between 10 nanoJy and 10 mJy. The former
value is of the order of the expected sensitivity for a 100 hour
observation with the nominal sensitivity ($1\sigma$) of the full
Square Kilometre Array ($A_{\mathrm{eff}}/T_{\mathrm{sys}}=20000$
m$^2$ K$^{-1}$, 350 MHz bandwidth). If NGC 253 is a prototype SBG
then the galaxy's radio flux of 50 nanoJy (that is
equivalent to $5\sigma$) will correspond to the flux from
the redshift of $z\simeq5$ while the galaxy's radio flux of
10 mJy will correspond to the flux from the distance of 100 Mpc
(that is $z=0.023$). We also checked that if the Milky Way is a
prototype normal galaxy then the galaxy's radio flux of 50
nanoJy will correspond to the flux from the redshift of $z\simeq2$,
while the galaxy's radio flux of 10 mJy will correspond to the flux
from the distance of 40 Mpc. We examined both the SBGs and normal
galaxies located at distances above 100 Mpc or 40 Mpc from the
Earth, respectively, including SFGs that are at the redshift of peak
star formation by means of the simulation
catalogue.

\subsection{Models}

\begin{figure*}
    \centering
    \includegraphics[width=0.75\textwidth]{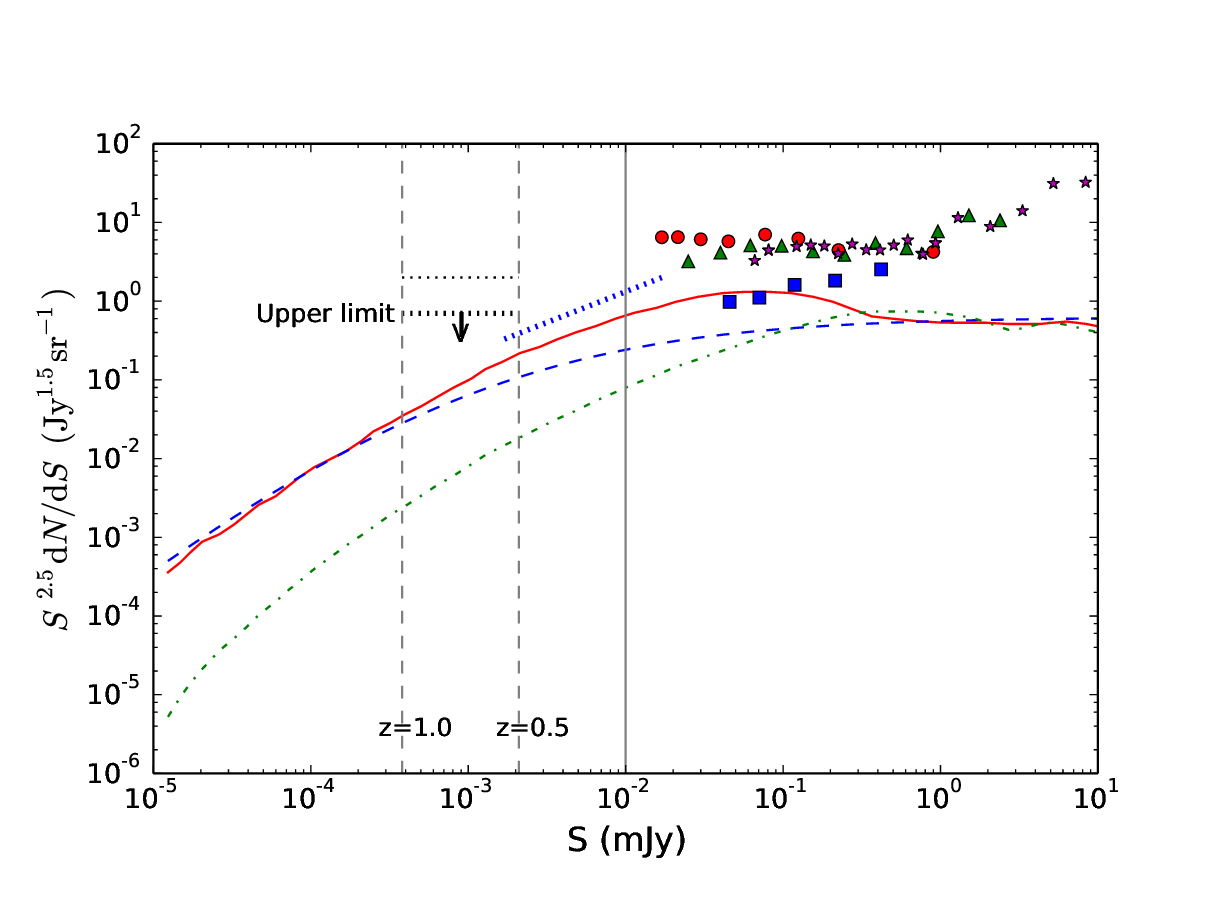}
    \caption{Source counts simulated by Wilman et al. (2008) for normal galaxies (solid line)
    and SBGs (dash-dotted line) including those at low radio fluxes, source counts observed in radio
    surveys at 1.4 GHz with the 10 $\mu$Jy flux density limit (a vertical solid line) shown by symbols
    (see Section 4.3), and the upper limit between $S=0.38-2.10$ $\mu$Jy derived in Section 4.4.}
    \label{Fig2}
\end{figure*}

We used the simulated $S^{2.5}\mathrm{d}N/\mathrm{d}S$ distribution
of radio source counts for normal galaxies and SBGs at 1.4 GHz taken
from Figure 4 of \citet[][]{Wilman2008} and normalised to
the Euclidean distribution
$\left(\mathrm{d}N/\mathrm{d}S\right)\propto S^{-2.5}$. To compute
the radio brightness (Jy sr$^{-1}$) we expressed the simulated
distributions in terms of $S\mathrm{d}N/\mathrm{d}S$ and integrated
these functions over flux density. We found that the computed radio
brightnesses at 1.4 GHz for SBGs and normal galaxies are 206 Jy
sr$^{-1}$ and 951 Jy sr$^{-1}$, respectively. 68 per cent of the
radio brightness generated by SBGs are from the sources with radio
flux of 0.01-1.0 mJy, while 74 per cent of the radio brightness
generated by normal galaxies are from the sources with radio flux of
1-100 $\mu$Jy. These large portions are consequences of the
fact that the $\mathrm{d}N/\mathrm{d}S$ distribution has different
slopes in the convergence region at the faint end,
$\left(S\mathrm{d}N/\mathrm{d}S\right)\propto S^{-0.5}$ and in the
Euclidean region at the bright end,
$\left(S\mathrm{d}N/\mathrm{d}S\right)\propto S^{-1.5}$
\citep[][]{Kellermann1987}, and the main contribution to the
radio brightnesses comes from the upper and lower bounds,
respectively. In the shell model \citep[][]{condon89} that
places all radio sources in a hollow, thin, spherical shell at
$z\simeq0.8$, we calculated the $\gamma$-ray fluxes (shown in Table
\ref{tab5}) expected from SFGs by using these radio brightnesses,
taking into account the coefficients of proportionality from Table
\ref{tab4}, and performing the K corrections
\citep[e.g.,][]{Peacock} for sources at this redshift. We found that
for various prototype starburst and normal galaxies the total
contribution of SFGs to the $\gamma$-ray background is between 10
per cent and 25 per cent. The contribution from the SBGs is about
two times higher than that from the normal galaxies. If the normal
galaxies are in the calorimetry regime at high redshifts (and have
the $\gamma$-ray-to-radio flux ratio equal to that of NGC 253) then
the total contribution of SFGs to the $\gamma$-ray background will
be between 30 per cent and 40 per cent.

The uniform model is an alternative to the shell model and
made on the reasonable assumption that the fainter sources are the
most distant ones. It assumes equal luminosities and that sources
are uniformly distributed in space with a number density,
$n(z)\propto(1+z)^3$, that decreases with the expansion of the
Universe. Taking the coefficients of proportionality from Table
\ref{tab4} and binning the radio brightnesses at 1.4 GHz in redshift
we estimated the contribution of SBGs and normal galaxies to the
$\gamma$-ray background. Given that SFGs are located at various
redshifts, we took the K corrections into account. Assuming that NGC
253 is a prototype SBG\footnote{We selected NGC 253 as a
prototype calorimeter, since its emission is less contaminated by an
active galactic nucleus than those from NGC 4945
\citep[][]{Wojaczynski2017}, NGC 3424, and UGC 11041
\citep[][]{Peng2019}.} we found the $\gamma$-ray background
produced by SBGs
is about fourteen times smaller than the observed one for
the \texttt{ULTRACLEANVETO} class in this energy range (Table
\ref{tab5}). If we consider the ultraluminous infrared galaxy, Arp
220, as a SBG prototype then the $\gamma$-ray background produced by
such sources will be about six times smaller than the observed
one. If we assume that the Andromeda galaxy (or the Milky
Way) is a prototype normal galaxy then the $\gamma$-ray signal
produced by such sources will be about 10 (3) per cent of the
observed $\gamma$-ray background flux. Thus, if NGC 253 and the
Milky galaxy are prototype starburst and normal galaxies, then we
can expect that the total contribution of SFGs to the $\gamma$-ray
background is about 7+3=10 per cent, while if we consider Arp 220
and the Andromeda galaxy prototype starburst and normal galaxies
then the contribution of SFGs will be about 17+7=23 per cent.
Assuming the ${\Lambda}CDM$ Universe and that the radio
luminosity of galaxies is of the order of the Milky Way's radio
luminosity, in the framework of the uniform model we can reproduce
two regions of the simulated $S^{2.5}\mathrm{d}N/\mathrm{d}S$
distribution for normal galaxies including the Euclidean region at 
the highest flux densities and the
convergence region in which the counts drop relative to the
Euclidean value. The observed differential source counts taken from
literature \citep[][and shown by circles, triangles, stars, and
squares, respectively]{Owen2008, Ibar2009, Bondi2008, Seymour2008},
the source counts simulated by \citet[][]{Wilman2008} for normal
galaxies (solid line) and SBGs (dash-dotted line), and those
obtained from the uniform model for normal galaxies in the
${\Lambda}CDM$ cosmology (dashed line) are shown in Figure
\ref{Fig2}. The results based on fluctuations due to weak confusing
sources (P(D) analysis by \citet[][]{Condon2012}) are shown by a
dotted line between 2 $\mu$Jy and 20 $\mu$Jy. The contribution of
active galactic nuclei is dominant above $1$ mJy, while star-forming
galaxies take over at fainter flux densities. It is worth noting
that the faint radio sky populations of low luminosity active
galactic nuclei cannot always be disentangled from normal galaxy
populations. If one assumes that the observed source counts
\citep[][]{Condon2012, Ibar2009} between 1-100 $\mu$Jy are totally
attributed to normal SFGs in the regime of CR proton escape and that
the Milky Way (Andromeda galaxy) is a prototype galaxy, then the
flux produced by these sources is 6\% (14\%) and 9\% (21\%) of the
total $\gamma$-ray background flux, respectively.

We found that the $\gamma$-ray background fluxes produced by
SFGs are similar in the two considered models, despite the facts
that in the shell model all the SFGs have different luminosities and
are located at $z=0.8$ and that in the alternative model
all the SFGs have equal luminosities and are distributed throughout space.

\subsection{Discussion}

The fact that the calculated flux from SFGs is smaller than
the unresolved $\gamma$-ray background flux suggests that

(i) other sources including blazars \citep[less than $20\%$
of the background,][]{anisotropies} and radio galaxies
\citep[][]{Hooper16} can contribute to the background at a significant
level;

(ii) more sophisticated models of galaxy evolution assuming
that the $\gamma$-ray-to-radio flux ratio is a function of
luminosity and that the luminosity distribution is in turn evolved
with redshift can be possible \citep[e.g.,][]{Fields2010, Linden2017};

(iii) the observational properties of low flux (distant)
SFGs can be significantly different from those of the nearby galaxies
detected in both the radio and $\gamma$-ray bands or a population of
low flux SFGs can be underestimated. In the latter case these low flux
SFGs are numerous and faint enough to dominate source counts below
the 10 $\mu$Jy limit of current radio surveys and extend to the 0.1
$\mu$Jy level.

Given that the possibilities (i) and (ii) are previously
considered in literature, below we discuss the low flux SFGs as
possible sources of the $\gamma$-ray background.

In the uniform model we checked and found that (a) if NGC 253 is
a prototype SBG then the SBGs with radio flux of 0.01-1 mJy, which produce a significant
fraction of the SBG contribution to the $\gamma$-ray background, are
located at redshifts $0.07<z<0.52$ and that (b) if the Andromeda
galaxy or the Milky Way are prototype sources then the normal
galaxies with radio flux of 1-100 $\mu$Jy which produce a
significant fraction of the normal galaxy contribution to the
$\gamma$-ray background are located at redshifts $0.05<z<0.42$ or
$0.09<z<0.66$, respectively. We found that the contributions of SBG
and normal galaxies at redshifts $1<z<2$ to the total $\gamma$-ray
flux produced by these types of sources, respectively, are 5 per
cent (SBGs such as NGC 253), 3 per cent (normal galaxies such as
the Andromeda galaxy) and 4 per cent (normal galaxies
such as the Milky Way). We considered the scenario
with a sharp change in the $\gamma$-ray-to-radio flux ratio above a
given redshift $z_{0}$ for normal galaxies.
We estimated the flux ratio at $z>z_{0}$ required to explain the unresolved
$\gamma$-ray background and found that the ratio should be by a factor of about
500-800 (depending on the energy bin), 300-500, and 100-200 for $z_{0}=$1.0, 0.75, and 0.5,
respectively, higher than that for the Milky Way.
Taking into account that the first two values exceed the $\gamma$-ray-to-radio
flux ratio for all the $\gamma$-ray detected SFGs by at least one order of magnitude,
this result suggests that the physical properties of galaxies significantly evolved
between the redshift of peak star formation and today.

To increase the background $\gamma$-ray flux from normal
galaxies located at high redshifts one can alternatively propose
that the numbers of such galaxies significantly exceeds that
simulated by \citet[][]{Wilman2008}. The 1.4-GHz flux densities of
Milky Way-like galaxies at $z=0.5-1.0$ are between $0.38$ $\mu$Jy
and $2.10$ $\mu$Jy and are below the sensitivity limit of previously
performed radio surveys. It is worth noting that the radio source
counts are based on interferometric images which are insensitive to
a smooth background, while the extragalactic radio background
brightness can have a significant contribution from a sufficiently
smooth distribution of faint radio sources
\citep[e.g.,][]{Condon2012}. Assuming the measured faint end index,
$\gamma=2.5$, of a power-law source count distribution,
$\mathrm{d}N/\mathrm{d}S\sim S^{-\gamma}$, from \citet{Owen2008}
(see Figure \ref{Fig2}) and Milky Way-type galaxies located between
$z\simeq0.5-1.0$ (with flux density between 0.35 $\mu$Jy and 2.1
$\mu$Jy), we computed the normalisation of the source counts at 1.4
GHz so that these source counts, \citep[e.g., Sect. 2.2
of][]{Singal} can account for the observed cosmic radio background
as reported by the ARCADE 2 collaboration \citep[][]{Fixsen}. We
found that the required number of faint sources significantly
exceeds that simulated by \citet[][]{Wilman2008}, see the (upper)
thin dotted line between two dashed vertical lines in Figure
\ref{Fig2}. We computed the background $\gamma$-ray flux expected
from this population of faint radio sources.
The $\gamma$-ray fluxes obtained assuming the $\gamma$-ray-to-radio flux ratio
for the Milky Way are $2.8\times10^{-8}$, $1.5\times10^{-8}$, and
$8.9\times10^{-9}$ GeV cm$^{-2}$s$^{-1}$ GeV$^{-1}$ sr$^{-1}$ at
energies of 3.5, 5.0, and 7.0 GeV, respectively. If the normal
galaxies at these high redshifts are CR calorimeters, one needs to
multiply these fluxes by a factor of 10 (adopting the
$\gamma$-ray-to-radio flux ratio for NGC 253) making the fluxes
two-three times higher than the total background $\gamma$-ray flux.
We obtained the upper limit on source counts shown in Figure \ref{Fig2}
corresponding to the level of source counts required to explain the
unresolved $\gamma$-ray background by the presence of low-flux SFG
calorimeters located at $z\simeq0.5-1.0$.
Next-generation, wide field, radio surveys will allow us to study
such populations of faint sources in the radio band and put tighter
limits on their number.

\section{Summary and outlook}

Astrophysical sources unresolved in $\gamma$ rays including SBGs and
normal galaxies can contribute to the extragalactic $\gamma$-ray
background. To evaluate the extragalactic background for testing its
possible discrete-source origin, one needs to subtract the
contribution of charged particles misclassified as photons and
contaminating the background. Faint unresolved sources are numerous
and present around the whole sky complicating the separation of
their $\gamma$-ray signal from the charged particle background.
Therefore, the filtration of charged particles is a prerequisite
procedure for classifying registered events on the bases of the
responses of detectors. The angular resolution of \textit{Fermi}-LAT
is significantly lower than those of radio and infrared instruments
and is limited at energies below 10 GeV by multiple scattering and
ultimately by the ratio of the silicon-strip pitch to silicon-layer
spacing. The advent of \textit{Fermi}-LAT allowed the detection of
$\gamma$ rays from nearby SBGs and the Andromeda galaxy broadening
our knowledge on the high energy emission from SFGs, while the
studies of SFGs with radio fluxes 1000 times lower than those of
these nearby SBGs or that of the Andromeda galaxy are routinely performed
with radio telescopes. This fact shows that a multi-frequency approach for
estimating the contribution of astrophysical source populations to
the $\gamma$-ray background is accessible.

We found evidence for both the geomagnetic shielding and the
east-west effect for $\gamma$-ray-like events of the \texttt{SOURCE}
and \texttt{CLEAN} classes in the \textit{Fermi}-LAT Pass 8 photon
data. To establish this evidence we binned $\gamma$-ray-like events
detected by \textit{Fermi}-LAT into bands both in the McIlwain L
parameter and energy. We found that the excesses in background
fluxes are present in the bands for which the geomagnetic shielding
is inefficient and checked that the mean probabilities of events in
these bands to be $\gamma$ rays are lower than those in the other
bands. We found that the fluxes for the \texttt{SOURCE}
(\texttt{CLEAN}) event class and also for the \texttt{CLEAN} event
class in the energy bands of 4-6 GeV and 6-8 GeV for large values of
McIlwain L-parameter are by a factor of about 2.5-4.0 (2.3-3.0)
higher than the corresponding fluxes for the lowest McIlwain L
bin. This finding shows that the filtration of charged particles
is possible by means of discarding the events in the 4-6 GeV and 6-8
GeV energy bands recorded during the time intervals when the
\textit{Fermi} spacecraft's geomagnetic position corresponded to the
high McIlwain L-parameter values. We also considered the possibility
to introduce a new event class in which events are selected not only
on the basis of the detector responses, but also regarding the
geomagnetic shielding effect. While our manuscript was in
preparation, the \textit{Fermi}-LAT collaboration uploaded a new
release Pass 8 data (P8R3) and showed that the P8R2 data are
contaminated by charged particles \citep[][]{Bruel2018}.
Thus, both the analyses conclude on the tendency that
charged particles are present in the P8R2 \texttt{SOURCE} class
though obtained by different methods given that their analysis
includes the signals from the ACD ribbons as observables.

We analysed the contribution of SFGs to the unresolved $\gamma$-ray
background at energies of 3 GeV to 8 GeV. At these energies of
$\gamma$ rays, the Universe is transparent and SFGs are expected to
significantly contribute to the $\gamma$-ray background. For this
analysis we used the simulation catalogue of radio sources
\citep[][]{Wilman2008} and calculated the $\gamma$-ray-to-radio flux
ratio on the basis of $\gamma$-ray and radio observations
of the nearby SFGs. We found that the total contribution of SFGs to
the $\gamma$-ray background is about 10-23 per cent depending on the
choice of prototype galaxies. It supports the conclusion by
\citet[][]{Linden2017} that multiple source classes provide
non-negligible contributions to the unresolved $\gamma$-ray
background. Under the assumption that the remaining flux can be
produced by normal Milky Way-type galaxies at high redshift emitting
in the calorimeter regime, we evaluated the required astrophysical
conditions and found that the normal galaxies at $z=1-2$ with radio
fluxes below 1 $\mu$Jy can produce a non-negligible contribution
only if evolutionary effects are present or alternatively the
$\mu$Jy radio population of SFGs is larger than simulated.
The latter possibility is of interest in consideration of
next generation radio surveys. We derived the upper limit on the
number of low-flux SFG calorimeters at $z=0.5-1.0$,
$\mathrm{d}N/\mathrm{d}S\leq0.7 S^{-2.5}$ Jy$^{1.5}$ sr$^{-1}$,
(shown in Figure \ref{Fig2}) corresponding to the level of
source counts required to explain the unresolved $\gamma$-ray
background by their presence. The full Square Kilometre Array with a
100-hour observation is expected to be sensitive to detect such
Milky Way-type galaxies to a redshift of $z\simeq2$.

\begin{figure}
    \centering
    \includegraphics[width=0.5\textwidth]{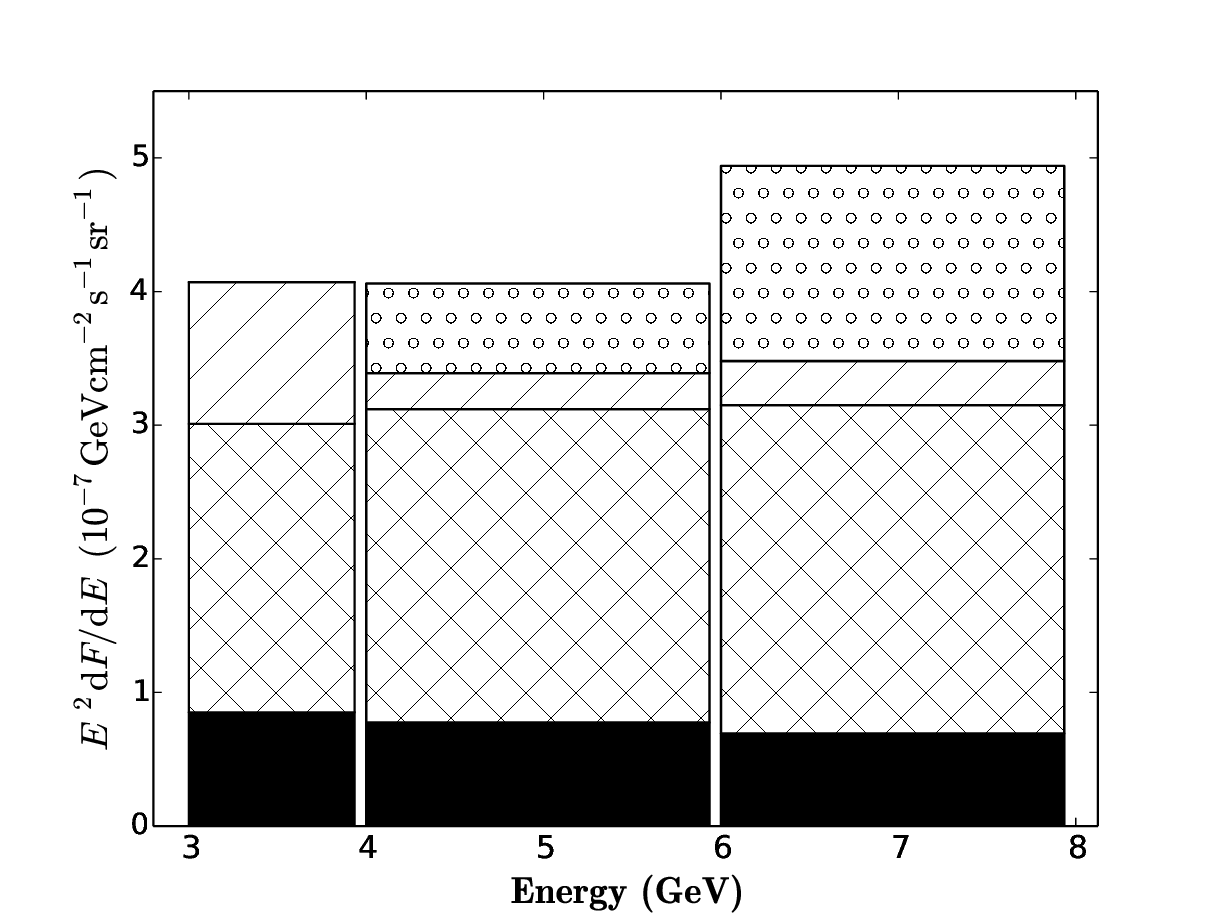}
    \caption{Background flux levels over the three energy bands.
    The total flux predicted for the populations of SFGs (regions filled with a solid color),
    the background flux level for the \texttt{ULTRACLEANVETO} class above the SFG-produced
    flux (cross-hatched regions), the difference between the \texttt{ULTRACLEANVETO} background flux
    and that for the filtered \texttt{SOURCE} class data (diagonal hatched regions),
    and the contribution of charged particles to the unfiltered \texttt{SOURCE} flux level
    (regions filled with circles).}
    \label{Fig3}
\end{figure}

Figure \ref{Fig3} illustrates the key results reported in the
paper including the filtered and unfiltered background flux levels
for the \texttt{SOURCE} class, the background flux level for the
\texttt{ULTRACLEANVETO} class, and the flux level resulting from the
populations of SFGs if Arp 220 and the Andromeda galaxy are considered
as prototypes.

\section{Acknowledgements}
We are grateful to the referee for the constructive suggestions that helped us
to improve the manuscript. DAP acknowledges support from the DST/NRF SKA post-graduate
bursary initiative.

\appendix

\section{}

Table \ref{tab2} shows the azimuthal flux ratio of
\textit{Fermi}-LAT events coming from the east to those coming from
the west in the the $45^{\circ}$ intervals normalised to that in the
McIlwain L bin, 0.980-1.128. Table \ref{tab3} shows the WP8CTCalTkrProb mean
probability which is the lowest in the McIlwain L bins with the excesses
in background fluxes. Table \ref{tab4} shows
the coefficient of proportionality between the radio flux of 1 Jy
and the $\gamma$-ray flux, $E\mathrm{d}F/\mathrm{d}E$, at energies
of 3.5 GeV, 5 GeV and 7 GeV in GeV cm$^{-2}$s$^{-1}$GeV$^{-1}$ for
each of the galaxies considered in this paper. Entries in bold and italic font
are explained in Section 3 of the main text.

\begin{table*}
\centering \caption{The azimuthal flux ratio normalised to that in
the McIlwain L bin, 0.98-1.12, for various McIlwain L bins and
energy bands obtained after masking events from the Galactic plane,
$-45^{\circ}<b<+45^{\circ}$ (or $-60^{\circ}<b<+60^{\circ}$).}
\begin{tabular}{ | c | c | c | c | c |}
\hline
& \multicolumn{4}{r}{Normalised azimuthal flux ratio} \\
\cline{3-5}
Event class & McIlwain L bin & 3-4 GeV & 4-6 GeV & 6-8 GeV\\
\hline
\texttt{SOURCE} & 1.128-1.276 & $0.98 (1.00)$ & $1.00 (0.99)$ & $1.10 (1.12)$ \\
\texttt{SOURCE} & 1.276-1.424 & $0.97 (0.97)$ & $1.02 (1.00)$ & $1.61 (1.67)$ \\
\texttt{SOURCE} & 1.424-1.572 & $0.99 (0.99)$ & $1.35 (1.39)$ & $\textbf{2.00 (2.17)}$ \\
\texttt{SOURCE} & 1.572-1.720 & $1.02 (0.98)$ & $\textbf{1.75 (1.92)}$ & $\textbf{1.61 (1.61)}$ \\
\hline
\texttt{CLEAN} & 1.128-1.276 & $0.97 (0.99)$ & $1.01 (0.99)$ & $1.05 (1.08)$ \\
\texttt{CLEAN} & 1.276-1.424 & $0.96 (0.96)$ & $1.00 (0.97)$ & $1.43 (1.47)$ \\
\texttt{CLEAN} & 1.424-1.572 & $0.98 (0.98)$ & $1.28 (1.30)$ & $\textbf{1.69 (1.85)}$ \\
\texttt{CLEAN} & 1.572-1.720 & $1.01 (0.98)$ & $\textbf{1.64 (1.82)}$ & $\textbf{1.47 (1.47)}$ \\
\hline
\texttt{ULTRACLEANVETO} & 1.128-1.276 & $0.95 (0.98)$ & $0.98 (0.99)$ & $1.01 (1.00)$ \\
\texttt{ULTRACLEANVETO} & 1.276-1.424 & $0.93 (0.94)$ & $0.96 (0.93)$ & $1.06 (1.08)$ \\
\texttt{ULTRACLEANVETO} & 1.424-1.572 & $0.97 (0.96)$ & $1.05 (1.08)$ & $\textit{1.05 (1.05)}$ \\
\texttt{ULTRACLEANVETO} & 1.572-1.720 & $0.97 (0.97)$ & $\textit{1.05 (1.20)}$ & $\textit{0.99 (0.87)}$\\
\hline
\end{tabular}
\label{tab2}
\end{table*}

\begin{table*}
\centering \caption{The probabilities, WP8CTCalTkrProb, for various
McIlwain L bins and energy bands obtained after masking events from
the Galactic plane, $-45^{\circ}<b<+45^{\circ}$ (or
$-60^{\circ}<b<+60^{\circ}$).}
\begin{tabular}{ | c | c | c | c | c |}
\hline
& \multicolumn{4}{r}{WP8CTCalTkrProb probabilities $\%$} \\
\cline{3-5}
Event class & McIlwain L bin & 3-4 GeV & 4-6 GeV & 6-8 GeV\\
\hline
\texttt{SOURCE} & 0.980-1.128 & $96.97 (96.85)$ & $97.29 (97.29)$ & $97.42 (97.32)$ \\
\texttt{SOURCE} & 1.128-1.276 & $97.64 (97.50)$ & $97.56 (97.36)$ & $96.73 (96.64)$ \\
\texttt{SOURCE} & 1.276-1.424 & $97.58 (97.41)$ & $97.21 (97.14)$ & $93.72 (93.50)$ \\
\texttt{SOURCE} & 1.424-1.572 & $97.44 (97.31)$ & $95.74 (95.57)$ & $\textbf{91.16 (90.84)}$ \\
\texttt{SOURCE} & 1.572-1.720 & $97.31 (97.06)$ & $\textbf{93.36 (92.73)}$ & $\textbf{89.78 (89.59)}$ \\
\hline
\texttt{CLEAN} & 0.980-1.128 & $97.22 (97.13)$ & $97.79 (97.74)$ & $98.22 (98.10)$ \\
\texttt{CLEAN} & 1.128-1.276 & $97.81 (97.65)$ & $98.02 (97.89)$ & $97.79 (97.69)$ \\
\texttt{CLEAN} & 1.276-1.424 & $97.78 (97.65)$ & $97.72 (97.70)$ & $96.11 (96.04)$ \\
\texttt{CLEAN} & 1.424-1.572 & $97.67 (97.54)$ & $96.75 (96.63)$ & $\textbf{94.61 (94.30)}$ \\
\texttt{CLEAN} & 1.572-1.720 & $97.28 (97.41)$ & $\textbf{94.76 (94.28)}$ & $\textbf{93.22 (92.87)}$ \\
\hline
\texttt{ULTRACLEANVETO} & 0.980-1.128 & $99.85 (99.84)$ & $99.86 (99.86)$ & $99.87 (99.86)$ \\
\texttt{ULTRACLEANVETO} & 1.128-1.276 & $99.84 (99.82)$ & $99.86 (99.85)$ & $99.87 (99.85)$ \\
\texttt{ULTRACLEANVETO} & 1.276-1.424 & $99.82 (99.83)$ & $99.83 (99.82)$ & $99.86 (99.83)$ \\
\texttt{ULTRACLEANVETO} & 1.424-1.572 & $99.79 (99.79)$ & $99.83 (99.82)$ & $\textit{99.82 (99.84)}$ \\
\texttt{ULTRACLEANVETO} & 1.572-1.720 & $99.79 (99.79)$ & $\textit{99.81 (99.78)}$ & $\textit{99.78 (99.85)}$\\
\hline
\end{tabular}
\label{tab3}
\end{table*}

\begin{table*}
\begin{minipage}{0.99\textwidth}
\centering \caption{The coefficient of proportionality between the
radio flux of 1 Jy and the $\gamma$-ray flux,
$E\mathrm{d}F/\mathrm{d}E$, computed for SFGs}
\begin{tabular}{ | c | c | c | c | c | c |}
\hline
SFG & Distance & Radio flux & Coeff. at 3.5 GeV & Coeff. at 5 GeV & Coeff. at 7 GeV\\
 & (Mpc) & at 1.4 GHz (Jy) &  GeV cm$^{-2}$s$^{-1}$GeV$^{-1}$Jy$^{-1}$ & GeV cm$^{-2}$s$^{-1}$GeV$^{-1}$Jy$^{-1}$ & GeV cm$^{-2}$s$^{-1}$GeV$^{-1}$Jy$^{-1}$ \\
\hline
NGC 253 & 3.3 & 5.9 & $3.8\times10^{-11}$ & $2.5\times10^{-11}$ & $1.7\times10^{-11}$ \\
M82 & 3.9 & 7.2 & $2.4\times10^{-11}$ & $1.6\times10^{-11}$ & $1.0\times10^{-11}$ \\
NGC 1068 & 15 & 7.5 & $1.5\times10^{-11}$ & $9.6\times10^{-12}$ & $6.2\times10^{-12}$ \\
Arp 220 & 72 & 0.3 & $1.6\times10^{-10}$ & $1.1\times10^{-10}$ & $7.2\times10^{-11}$ \\
Andromeda & 0.78 & 8.6 & $1.1\times10^{-11}$ & $6.8\times10^{-12}$ & $4.2\times10^{-12}$ \\
Milky Way & -- & 16.8$^{\dagger}$ & $4.8\times10^{-12}$ & $2.6\times10^{-12}$ & $1.5\times10^{-12}$ \\
\hline
\end{tabular}
\footnotetext{$^{\dagger}$ at the distance of 1 Mpc from the Milky
Way} \label{tab4}
\end{minipage}
\end{table*}

\end{document}